\documentclass[mti,article,accept,moreauthors,pdftex,10pt,a4paper]{mdpi}

\usepackage{graphicx}
\usepackage{verbatim}
\usepackage[english]{babel}
\usepackage[utf8]{inputenc}
\usepackage[T1]{fontenc}
\usepackage{algorithmic}
\usepackage{algorithm}
\usepackage{time}
\usepackage{array}
\usepackage{multirow}
\usepackage{url}

\firstpage{1}
\articlenumber{23}
\doinum{10.3390/mti2020023}
\pubvolume{2}
\pubyear{2018}
\copyrightyear{2018}
\history{Received: 10 March 2018; Accepted: 3 May 2018; Published: 7 May 2018}

\Title{A Study on the Use of Eye Tracking to Adapt Gameplay and Procedural Content Generation in First-Person Shooter Games}

\Author{{Jo\~ao Antunes} and Pedro Santana *\href{https://orcid.org/0000-0002-4357-1546}{\orcidicon}}
\AuthorNames{Jo\~ao Antunes and Pedro Santana}

\address [1]{ISCTE-Instituto Universit\'ario de Lisboa (ISCTE-IUL) and Instituto de Telecomunica\c c\~oes (IT), {\mbox{1649-026} {Lisboa, Portugal}; jmmea@iscte-iul.pt}}

\corres{\hangafter=1 \hangindent=1.05em \hspace{-0.82em}Correspondence: pedro.santana@iscte-iul.pt; Tel.: +351-217-650-558}

\abstract{This paper studies the use of eye tracking in a First-Person Shooter (FPS) game as a~mechanism to: (1) control the attention of the player's avatar according to the attention deployed by the player; and (2) guide the gameplay and game's procedural content generation, accordingly. This results in a more natural use of eye tracking in comparison to a use in which the eye tracker directly substitutes control input devices, such as gamepads. The study was conducted on a custom endless runner FPS, Zombie Runner, using an affordable eye tracker. Evaluation sessions showed that the proposed use of eye tracking provides a more challenging and immersive experience to the player, when compared to its absence. However, a strong correlation between eye tracker calibration problems and player's overall experience was found. This means that eye tracking technology still needs to evolve but also means that once technology gets mature enough players are expected to benefit greatly from the inclusion of eye tracking in their gaming experience.}


\begin{document}

\section{Introduction}
\label{chapter:introduction}

With eye tracking dating back from the XVIII century \cite{javal_1879}, many have been the applications for this technology, such as medicine \cite{holzman_proctor_hughes_1973}), robotics \cite{mcmullen_hotson_2014,gomes2016gaze}), advertising  \cite{krugman_fox_fletcher_fischer_rojas_1994}) and, more recently, computer~games~\cite{smith_graham_2006}.~In the past decade, research on computer games tried to compare traditional input (e.g., mouse, keyboard, and gamepad) with eye tracking input, in terms of action accuracy and \mbox{responsiveness \cite{leyba_malcolm_2004,isokoski_martin_2006,smith_graham_2006,isokoski_joos_spakov_martin_2009,dorr_pomarjanschi_barth_2009}.} These comparisons were often made after asking players to compete against each other or asking users to complete a given task, using the different input methods. Overall, these comparisons provided mixed results, with some studies claiming that the use of eye tracking contributed to better task completion \cite{dorr_pomarjanschi_barth_2009}, while others claiming that traditional input devices provided better overall results \cite{leyba_malcolm_2004}. Hence, these previous studies have provided contradictory results regarding the effectiveness of the use of eye tracking in computer games as a direct control input, which may be an indication that eye tracking is not best suited to direct input control.

Bearing the limitations of eye tracking as a simple direct control input in mind, in this paper, we~propose to use it  to control the attention of the player's avatar and the game's procedural content generation. This use of eye tracking is focused in mapping the player's and avatar's attention processes, which we believe to be much more natural and useful than allowing the player to directly control a pointer with the eyes, which has no actual mapping to real life. That is, instead of replacing the traditional input for an eye tracker, we are more interested in studying meaningful ways about how eye tracking can improve gameplay, which includes co-existence with traditional inputs. This~also means that, instead of analysing objective performance-based data, as in previous studies, we are more interested in a subjective analysis of how eye tracking improves enjoyability and how well the player adapts to the technology.

\begin{figure}[H]
\centering
\includegraphics[width=12cm]{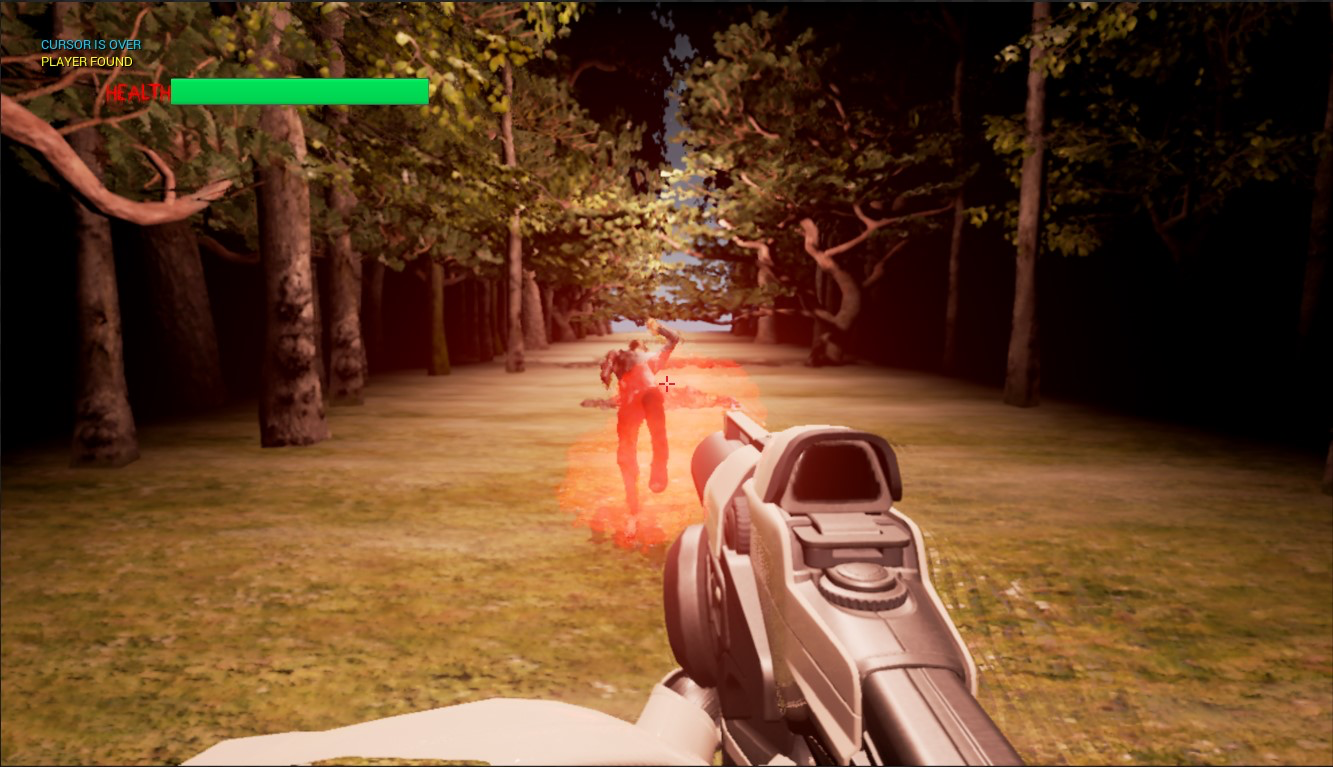}

\caption{A zombie being killed in Zombie Runner.\vspace{-6pt}}
\label{fig:zombiekilled}
\end{figure}

To demonstrate the value of the herein proposed alternative uses of eye tracking in computer games, we developed and tested our own endless runner First-Person Shooter (FPS), Zombie Runner (see Figure~\ref{fig:zombiekilled}). In this game, shot accuracy, automatic obstacle avoidance, and procedural obstacle spawn probability are controlled as a function of the avatar's attention model, which, in turn, operates~according to the player's gaze estimated with an affordable eye tracker. The goal is to better represent the player's actions in the game, thus contributing to a more immersive experience. Based on a set of testing sessions, we conclude that the use of eye tracking provides a more challenging and immersive experience to the player. Participants reported better levels of satisfaction while playing the game with the gaze tracking turned on. However, a strong correlation between eye tracker calibration problems and player's overall experience was found. This means that eye tracking technology still needs to evolve but also means that once technology gets mature enough players are expected to benefit greatly from the inclusion of eye tracking in their gaming experience.

This article is an extended and improved version of a poster paper \cite{antunes2017} and it is organised as follows. Section~\ref{chapter:leteraturesurvey} presents an overview of previous and related work. Then, in Section~\ref{chapter:developmentimplementation}, Zombie~Runner is described and its implementation detailed. Section~\ref{chapter:evaluationdiscussion} describes the experimental setup and analysis the obtained results. Finally, Section~\ref{chapter:conclusionsfuturework} presents some conclusions and provides some future work directions.

\section{Related Work}
\label{chapter:leteraturesurvey}

Eye tracking is the process of estimating one's gaze direction, identifying the object in which the the subject is focused \cite{lukander_2004,arai_mardiyanto_2011}. Eye tracking dates from the XVIII century, in which persistent images to describe the human eye movements were used \cite{wells_1792}. In the XX century, the first eye movement measures through a non-intrusive method using photographs and light reflections were made \cite{dodge_cline_1901}. In~the 1980s, with the evolution of computing capacity, it became possible to perform real-time eye tracking with access to video, what opened the possibility of human--machine interaction \cite{singh_singh_2012}. With~increasingly more accessible prices \cite{shell_vertegaal_cheng_skaburskis_sohn_stewart_aoudeh_dickie_2004,smith_vertegaal_sohn_2005}, the use of eye trackers increased in several areas, such as marketing \cite{krugman_fox_fletcher_fischer_rojas_1994}, psychology \cite{holzman_proctor_hughes_1973} and, more recently, computer games \cite{smith_graham_2006}.

Eye tracking has been explored in computer games as an alternative to traditional input methods, such as mouse or keyboard \cite{smith_graham_2006,isokoski_joos_spakov_martin_2009}. By testing gaze input versus mouse input in three different computer games, Smith and Graham \cite{smith_graham_2006} concluded that the use of eye tracking can provide a more immersive experience to the player. Isokoski and Martin \cite{isokoski_martin_2006} conducted a preliminary study on the use of eye tracker in First Person Shooters. Each participant in the test was asked to play the same game using three different input method schemes: (1) mouse, keyboard, and eye tracker; (2) only mouse and keyboard; or (3) a console gamepad. The conclusions were not exactly encouraging, suggesting that the performance with the eye tracker was quite inferior to the other two. However, Isokoski and Martin attributed these results to the players' greater knowledge and contact with the traditional input methods, suggesting that this scenario could change with more training.

Other studies achieved similar conclusions. Leyba and Malcolm \cite{leyba_malcolm_2004} created a simple test in which the player was asked to eliminate twenty-five balls that moved around the screen at different velocities. The player would move the pointer using the mouse or the eye tracker and would eliminate the balls by clicking on them with the mouse. Two conditions were tested: with and without time limit to complete the task. The results showed that, without time limit, precision and time to complete the task was worse while using the eye tracker than when using a mouse. The same results were obtained for the no time limit condition, in which performance was based on percentage of balls eliminated by the player. Michael Dorr et al. \cite{dorr_pomarjanschi_barth_2009} achieved totally opposite results. After creating and adapting a clone of the classic game Breakout, twenty players were asked to participate in a tournament. Players were separated in pairs. The two players of each pair played against each other, one using the mouse and the other using an eye tracker. The control inputs were swapped between rounds. The results showed that the players who used the eye tracker achieved higher scores and won more rounds. The players also stated that using the eye tracker was highly enjoyable. These discrepant results between studies suggest that the type of game and development method of the same game are key elements to achieve a satisfying final result.

Bearing the limitations of using eye tracking as a simple direct control input in mind, we propose to use it to control the attention of the player's avatar and the game's procedural content generation. This use of eye tracking is focused in mapping the mental state of the player and her/his avatar, which~we believe to be much more natural and useful than controlling a pointer with the eyes, which~has no mapping to real life. Another alternative and interesting use of eye tracking, not tackled in this paper, is to know when to actively redirect the player's attention \cite{perreira_2007}.

Procedural Content Generation (PCG) concerns all creation of game content (e.g., sounds, levels, objects, characters, and textures) with algorithms, with limited or indirect stimuli from the user \cite{shaker_togelius_nelson_2014}. PCG is a way of coping with the daunting task of manually creating and populating massive open worlds. Besides this more traditional use of PCG, some researchers proposed that, through the analysis of the interaction between the player and the game, PCG could be used to create a playing experience that adapts itself to the player \cite{browne_yannakakis_colton_2012,yannakakis_2012,gow_baumgarten_cairns_colton_miller_2012}, improving the game's replay value. This new approach is known as Experience-Driven Procedural Content Generation (EDPCG) \cite{yannakakis_togelius_2011}.

Within the EDPCG framework, it is possible to generate levels that are adapted to the strengths and limitations of the player, in an attempt to maximize the fun factor. Many studies propose that the fun and challenge factors are directly linked \cite{iida_takeshita_yoshimura_2003,spronck_sprinkhuizen-kuyper_postma_2004,andrade_ramalho_santana_corruble_2005,yannakakis_hallam_2007,olesen_yannakakis_hallam_2008,lankveld_spronck_herik_rauterberg_2010,sorenson_pasquier_2010}, which means that, to tune the fun factor, one~often has to tune the challenge factor.  In this line, in this paper, we propose the integration of PGC with gaze tracking so as to adapt the challenge the player as to face according to his/her form of playing and, \mbox{as a result}, to improve the fun factor.

\section{Zombie Runner}
\label{chapter:developmentimplementation}

The game herein presented, Zombie Runner, was purposely designed to integrate gaze tracking at its core mechanic as a way of  estimating the player's attention and use those estimates to control the avatar's attention, that is, as a way of implementing a mapping between the player's and the avatar's attention processes. This is in contrast to the typical use of eye tracking to control avatar's motor actions (e.g., to aim the weapon). In Zombie Runner, motor actions are controlled via a traditional~input, a gamepad. This way, gaze and hand orthogonally and, thus, more naturally, control their virtual counterparts. To better analyse the advantages and disadvantages of the use of eye tracking in games, focus should be given to game genres where the player's visual attention has to be shared between different elements in the scene under tight temporal restrictions, where the eye tracking use is more central and challenging. For this reason, Zombie Runner has been designed and implemented as the common genre (of high impact) FPS. The character's running action is fully automated to reduce the number of actions the player had to memorize and control and, hence, reduce variability in the game evaluation phase.

\subsection{System Configuration}
\label{sec:system-config}

\begin{figure}[H]
\centering
\includegraphics[height=6cm]{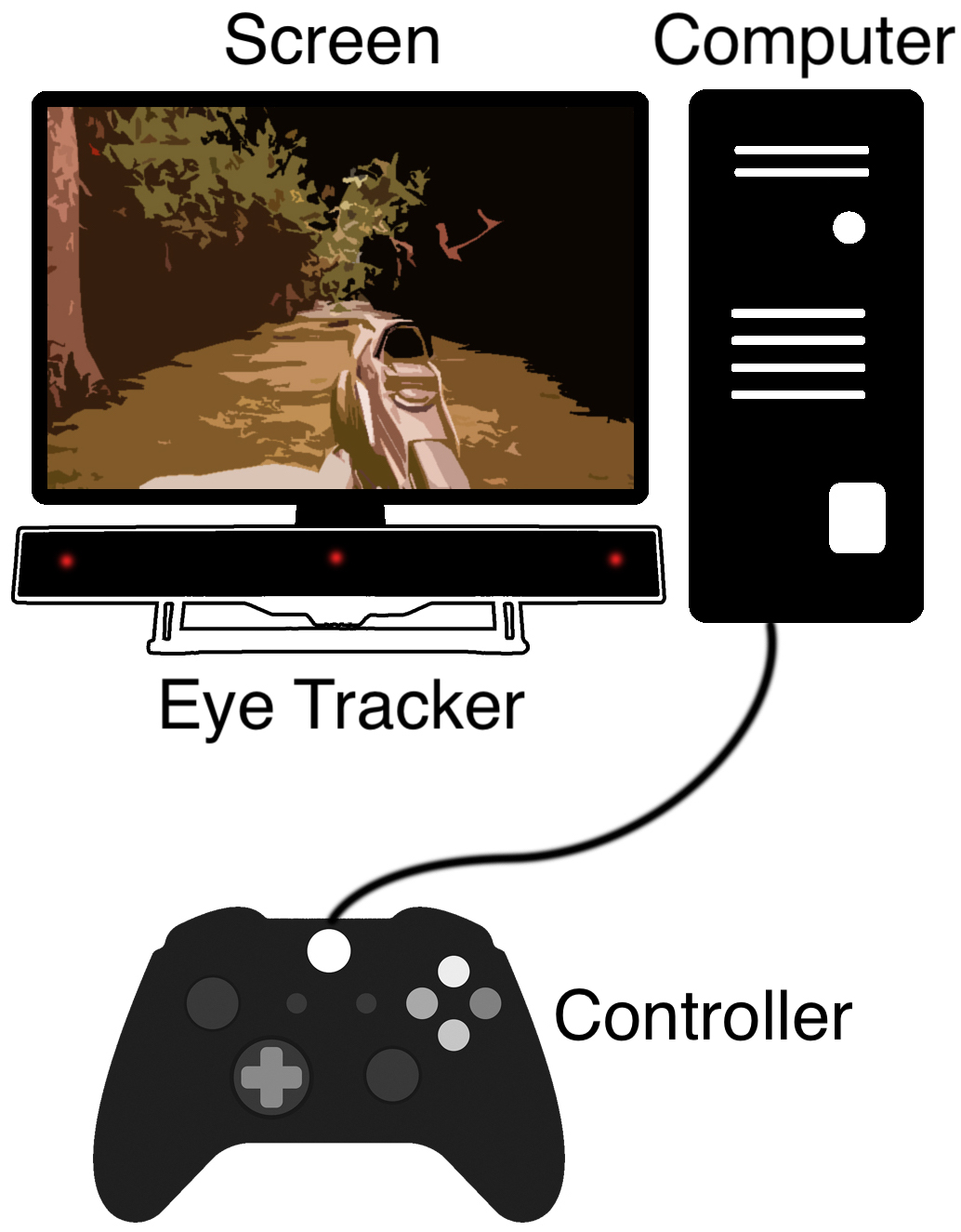}\hspace{0.2cm}\includegraphics[height=6cm]{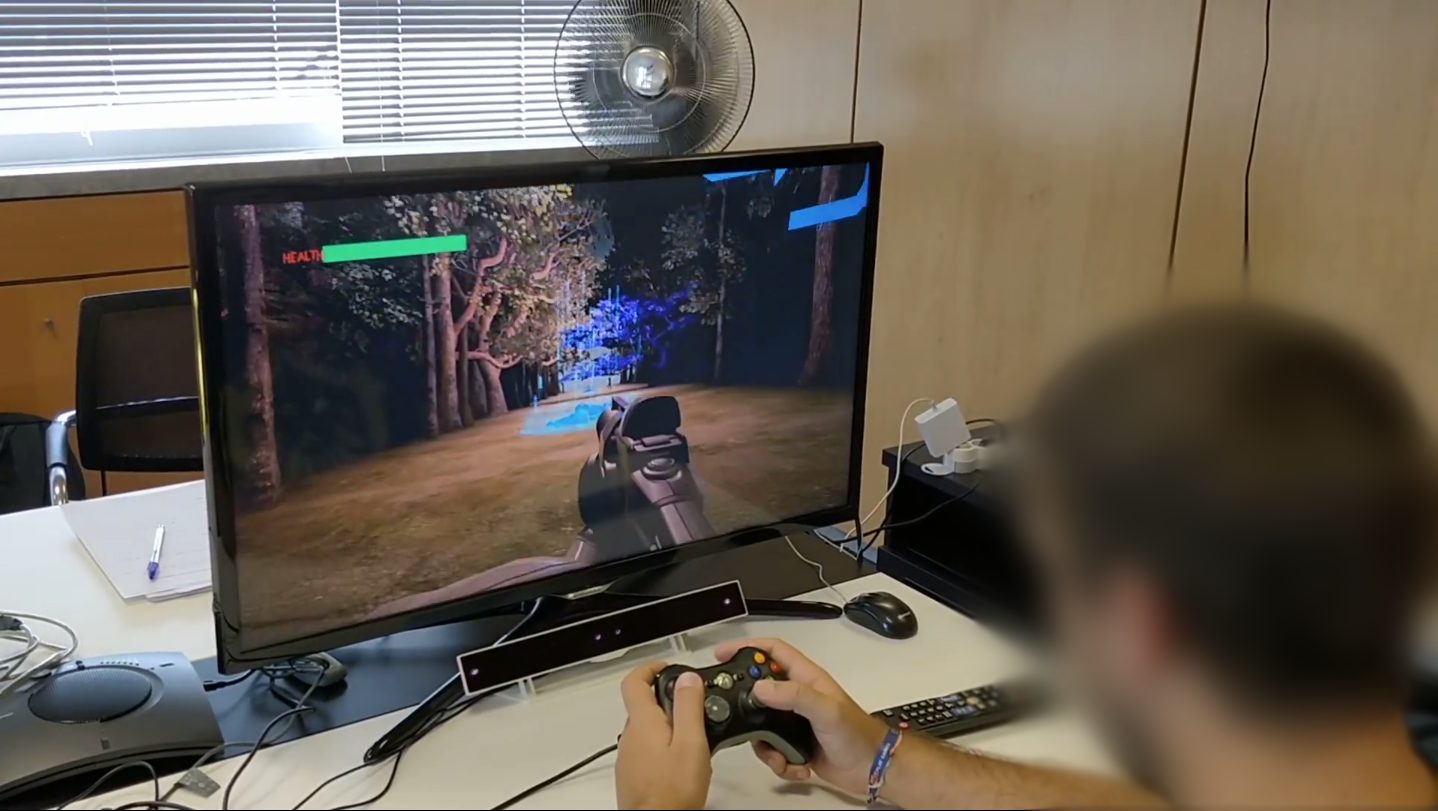}

\caption{The hardware configuration and a test subject during an evaluation session.\vspace{-6pt}}
\label{fig:testroomsetup}
\end{figure}

The game, developed in Unreal Engine,  has been with the following hardware configuration in mind (see Figure~\ref{fig:testroomsetup}): a low-cost Gazepoint GP3 eye tracker; a Microsoft Xbox gamepad; and~a 32~inch computer screen. The player sits in front of the computer screen, with the gamepad in hand. The~eye tracker sits below the computer screen. Behind it, a laptop running the game is available for the research team during the evaluation sessions.

In Zombie Runner, eye tracking is ensured by the Gazepoint's control software \cite{gazepoint_control}, which takes control of the mouse cursor to direct it towards the gaze point in the screen. Thus, Zombie Runner only needs to be sensitive to the mouse's cursor  to determine the player's gaze in screen coordinates. The~eye tracker's vendor claims a visual angle accuracy between $0.5^{\circ}$ and $1.0^{\circ}$ at 60 Hz, which we consider to be sufficient to assess  which in-game elements are being attended by the player. A recent study confirms the vendor's claimed accuracy \cite{zugal2014low}, provided that the user does not wear glasses, the~testing environment offers proper lighting conditions, and a correct calibration procedure is carried~out.

\begin{figure}[H]
\centering
\includegraphics[height=7cm]{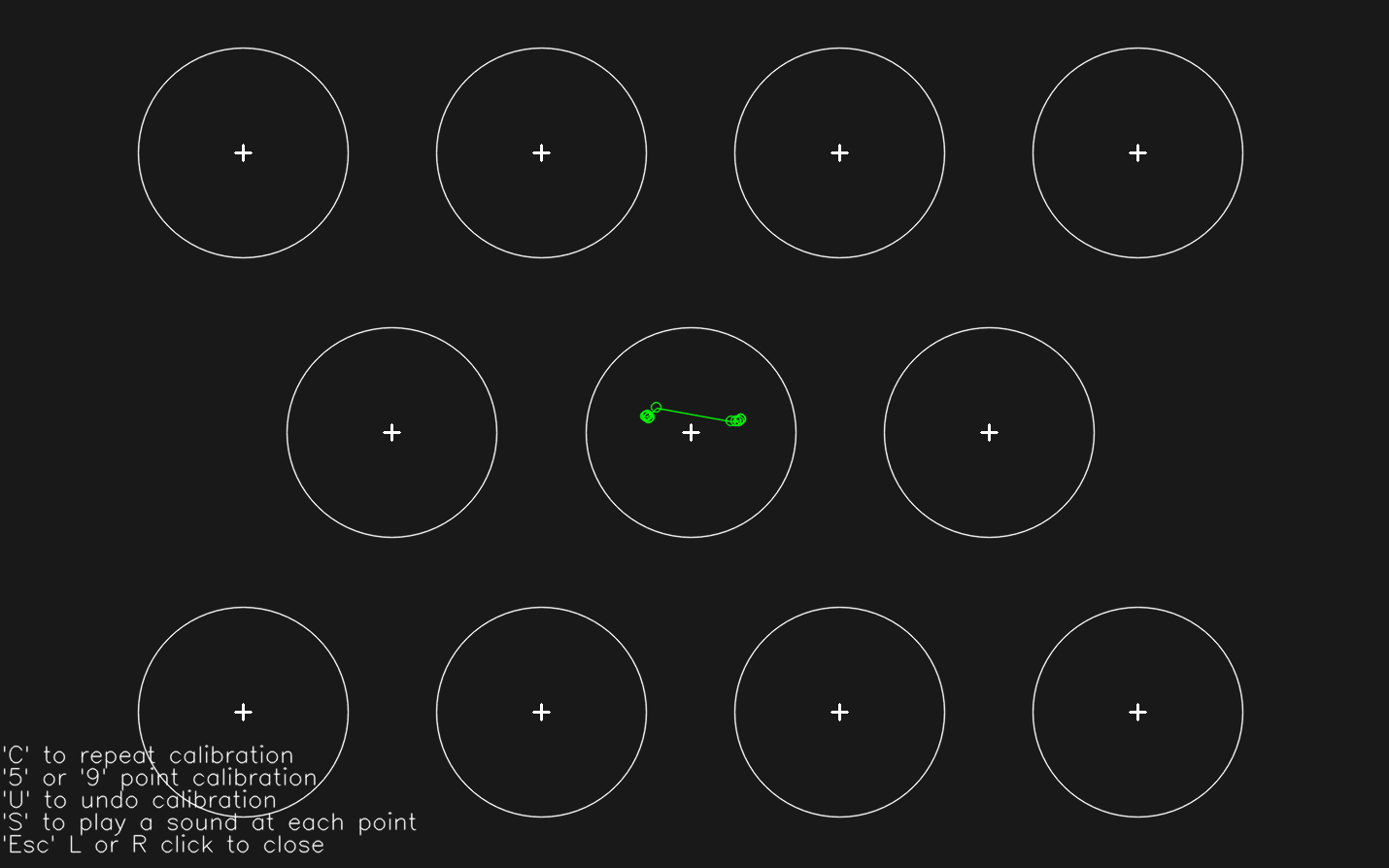}

\caption{The Gazepoint Control software's screen used to test the calibration results. The user is asked to look at the centre of each circle. The calibration is assumed to be successful if the gaze, represented~in green, never lands outside the circle being attended by the user.\vspace{-6pt}}
\label{fig:calibration-screen}
\end{figure}

To attain a correct calibration, Zombie Runner uses a nine-point calibration procedure shipped with the Gazepoint's control software. Before playing the game, the user must confirm that the calibration is correct by running a calibration validation procedure. This procedure consists of gazing at the centre of several circles displayed in the screen (see Figure~\ref{fig:calibration-screen}). This calibration is assumed to be successful ({good enough}) if the gaze never lands outside the circles being attended by the user.

\subsection{Game Rules and Mechanics}

Zombie Runner follows the following set of rules. The main objective of the game is to ensure that the avatar survives for as long as possible while running along a corridor with a non controllable constant forward movement. The player can achieve this by killing enemies and by \textit{noticing} elements in the scene (enemies and obstacles). The player can kill enemies by aiming and sooting the avatar's gun via the gamepad. The player can \textit{notice} the several elements in the scene by actively looking at them on the screen for a sufficient amount of time. The gaze of the player is estimated with the eye tracker. An element noticed by the player is also noticed by the avatar.

\begin{figure}[H]
\centering
\includegraphics[width=9.5cm]{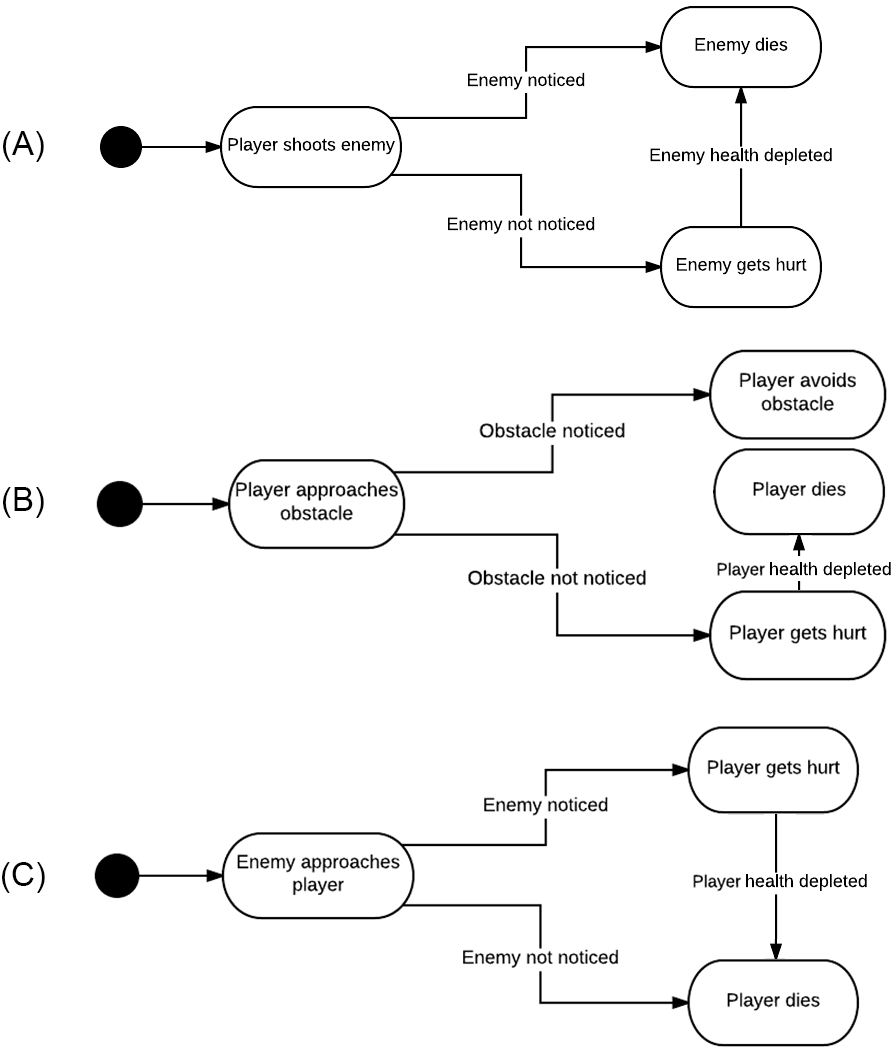}

\caption{State diagrams for: when the avatar shoots an enemy (\textbf{A}); when the avatar approaches an~obstacle (\textbf{B}); and when an enemy approaches the avatar (\textbf{C}). The arrows indicate state transitions and the text in them are the conditions necessary to trigger another state of the interaction.\vspace{-6pt}}
\label{fig:gameflowinteractions}
\end{figure}

If the avatar approaches a previously \textit{noticed} obstacle, it will automatically avoid that obstacle (i.e.,~jump over a rock or dodge a hanging tree branch). If this obstacle were not noticed early enough, it~will not be seen by the avatar, which will result in a collision and subsequent avatar's health decrement. A shot enemy that was noticed dies instantly, whereas it will only be hurt if unnoticed. This intends to simulate the lack of shot accuracy resulting from an insufficient focus on the enemy. If~a noticed enemy approaches the avatar, it will attack, causing the avatar to lose health. If the enemy was not noticed early enough, its attack will cause instant avatar's death. The idea is that the avatar is able to dodge the attack provided the attacker had been noticed. The avatar dies when its health reaches zero, leading to the game being over. Figure~\ref{fig:gameflowinteractions} depicts the state diagrams associated to these game rules.

\subsection{Game Implementation}
\label{sec:devunrealengine}

Zombie Runner was developed on Unreal Engine. The programming of Zombie Runner was split between C++ and Blueprints, the Unreal's visual scripting language. C++ was used for the core algorithms while Blueprints was used to program actor behaviour, such as enemies and obstacles. The~game is based on the Unreal's FPS template, which already implements the expected behaviour for a~generic game of the genre, including shooting, walking, and aiming mechanisms.

\begin{figure}[H]
\centering
\includegraphics[width=14cm]{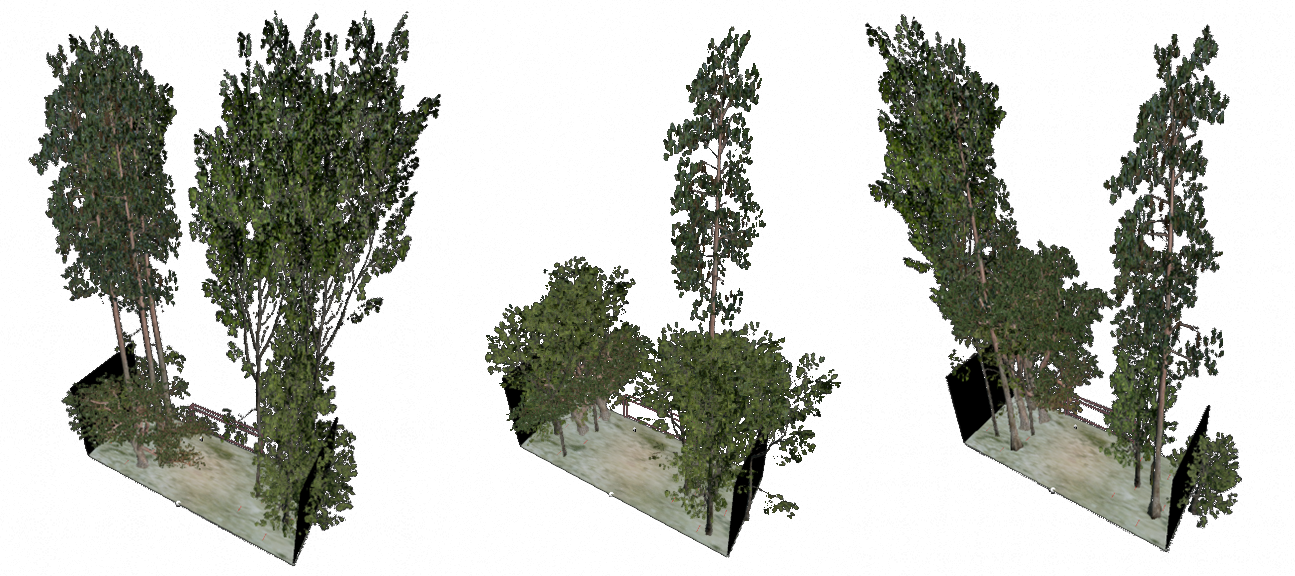}

\caption{Tiles populated with procedurally placed obstacles. Note that the tree branches in the central region of the tiles are implemented as small rotated trees.\vspace{-6pt}}
\label{fig:tilegeneration}
\end{figure}

The corridor in which the game takes place is made of an unidimensional array of 3D tiles (see~Figure~\ref{fig:tilegeneration}), which are procedurally generated and removed from the scene once the avatar passes by the them. Tiles are comprised of one plane for the floor, two planes for the side walls, an array of marker points distributed randomly on the sides of the floor (not in the region in which the avatar will be crossing), and an array of tree models. Every time a tile is spawned, the array of marker points is traversed. For each marker, there is a $66\%$ probability of spawning a tree (non-obstacle) in its position, plus a small random offset. This tree is spawned with a random rotation. The tree's height is also randomly set between reasonable values. The result are sufficiently different tiles that, when placed in succession, give the feeling of a dense and varied forest on each side of the player's sight. Figure~\ref{fig:tilegeneration} shows three different results of this algorithm applied to the tile generation.

Fifteen tiles are initially generated. After the eighth tile is spawned, obstacles or enemies can start being spawned with them. This value was achieved by trial and error and produces an initial tile generation that gives enough time for the player to settle in and prepare for the incoming obstacles and enemies. After this initial generation, all subsequent tile generation is  recursively and procedurally handled by the tiles themselves. An obstacle (i.e., a rock or an hanging tree branch) or an enemy has a $33\%$ probability of being procedurally spawn in each new tile. If it does occur, there is a $55\%$ probability for either spawning an obstacle or an enemy. A spawned enemy has a $20\%$ probability of being spawned as a \textit{runner}, which has double the speed of a \textit{walker}, the default behaviour of an enemy. These percentages have been tuned according to a set of informal tests to ensure that the game was playable without training.

\begin{figure}[H]
\centering
\includegraphics[width=7cm]{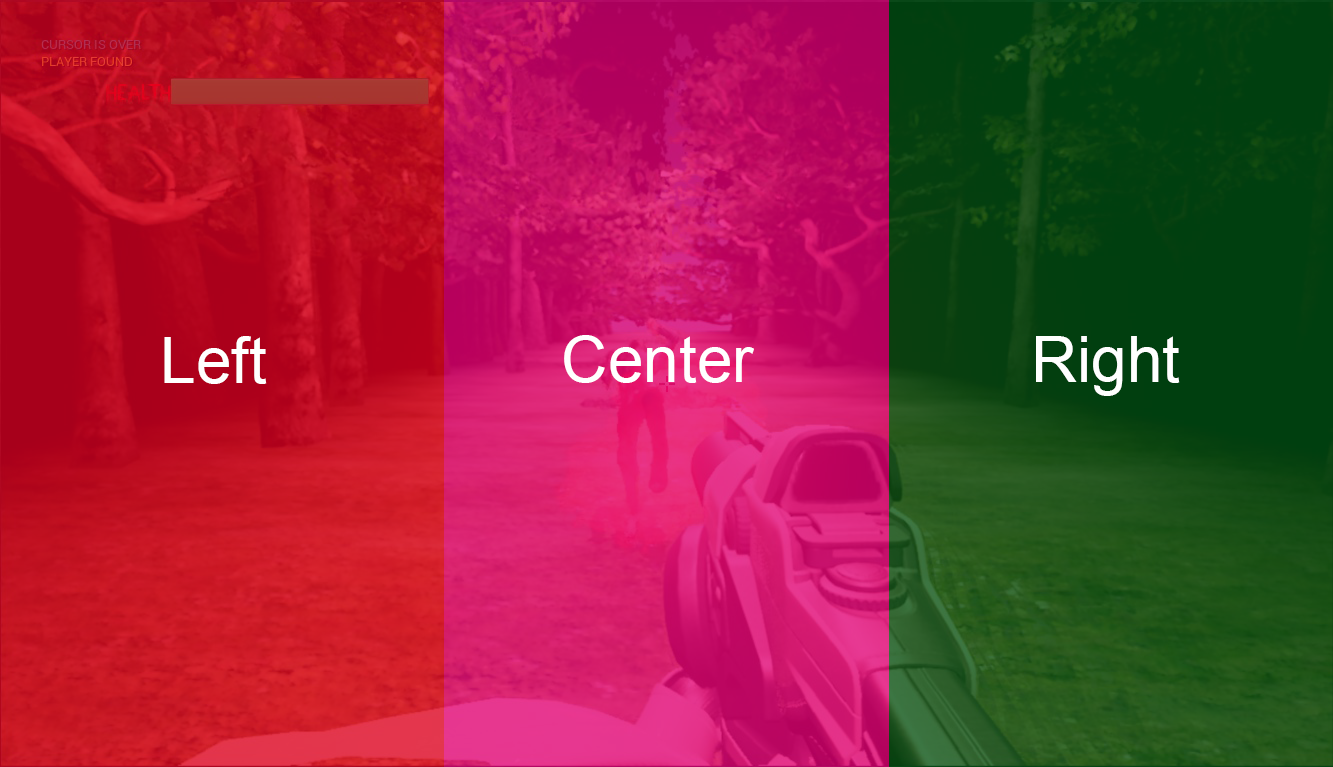}

\caption{Tile/screen regions used to determine where to spawn an obstacle or enemy.}
\label{fig:screensplit}
\end{figure}

An obstacle or an enemy can be spawned in three different regions in the tile (see Figure~\ref{fig:screensplit}): central~region, left region, or right region. If the previous spawned obstacle or enemy was spawned in either the left or the right regions, the new one will be spawned in the central region. This adds variety to the game and avoids objects of the same type to be spawned close to each other. If the previous spawned obstacle or enemy was spawned in the central region, the new one will be spawned in either left or right regions, depending on the player's gaze. Concretely, the obstacle or enemy is spawn in the region of the tile least gazed by the player during the time spent running over the two previous tiles. This~forces the attention of the player to be often alternating between regions, thus increasing the challenge. When the obstacle is spawned in the central region, its asset is a rock, being a tree branch when spawned on one of the side regions. Tree branches are implemented as laying small trees (see~Figure~\ref{fig:tilegeneration}).

Zombie Runner relies on ray casting to determine which \textit{object} (term hereafter used to generally refer to obstacles and enemies) present in the virtual scene  is being attended by the player, given~the gaze position in the screen. Assuming that the avatar's camera (the one rendering the images presented to the player) is located at world coordinates $\mathbf{o}$ and the player is gazing towards the screen's local coordinates $\mathbf{s}$, the system determines which object is being attended by the player by casting into the scene a parametric ray $\mathbf{r}(t)=\mathbf{o}+t(\Phi(\mathbf{s})-\mathbf{o})$, where the function $\Phi(\cdot)$ transforms a point from screen coordinates to world coordinates, given the virtual camera's field of view and pose. The closest intersected object is taken as the one being attended by the player.

Due to the rapid nature of FPS games (which induces frequent gaze shifts) and the limited accuracy of low-cost eye tracking technology, demanding for gaze fixations to occur with high certainty (i.e., to lock in a given object for a significant amount of time) before accepting that a given object was attended by the player could result in many false negatives. Here, a false negative means penalizing the player because a given object was not attended when, in practise, the player feels that the object was attended (even if only momentarily). These events are more harmful to the player's engagement level than the other way around, that is, to erroneously assume that the player has attended to a not actually attended object. Hence, the approach followed in Zombie Runner is to label a given object as noticed if the gaze of the player intersects that same object over an \textit{accumulated} (i.e., time is not reset when gaze leaves the object) period of 0.5 s. This period was obtained from a set of informal tests. The~approach of considering accumulated time ensures that an object is marked as attended/noticed even if the player frequently gazes across several objects or the eye tracker estimates jitter around the player's gaze, meaning it frequently lands off the attended object. Thus, this approach implements a~filtering process that trades-off false positives and false negatives in a way that suits the game's~needs.

\begin{figure}[H]
\centering
\includegraphics[width=10cm]{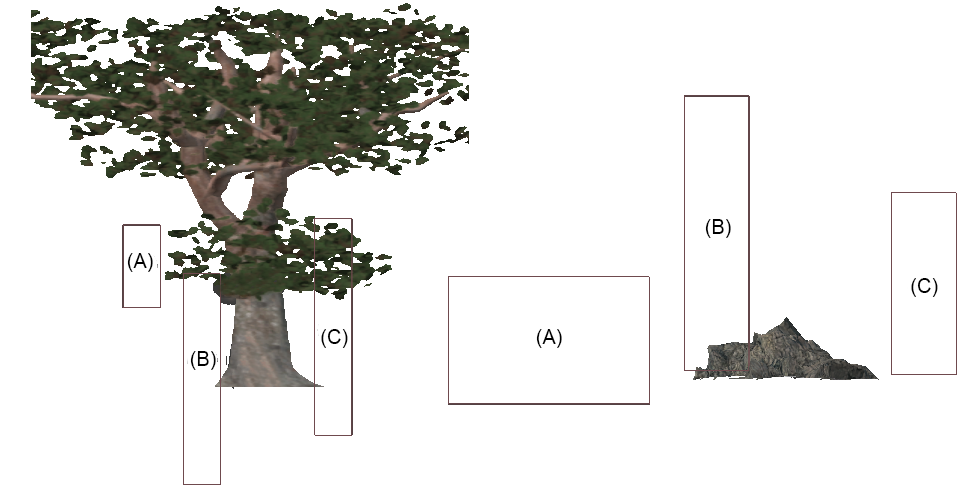}

\caption{A lateral view of the meshes composing the two types of obstacles and their associated three intersection bounding boxes, assuming that the avatar approaches the obstacles from the left. Boxes~labelled as (\textbf{A}) are used for detecting ray--obstacle intersections (to determine which object is being attended by the player). In the case of the tree, box (\textbf{A}) surrounds the tree's horizontal hanging branch, that is, the actual obstacle to the avatar (see Figure~\ref{fig:tilegeneration}). Boxes labelled as (\textbf{B}) are used to detect the moment the avatar is about to collide against the obstacle. Boxes labelled as (\textbf{C}) allow the system to detect the moment the avatar passes beyond the obstacle.}
\label{fig:boxesobstacles}
\end{figure}

\begin{figure}[H]
\centering
\includegraphics[width=9cm]{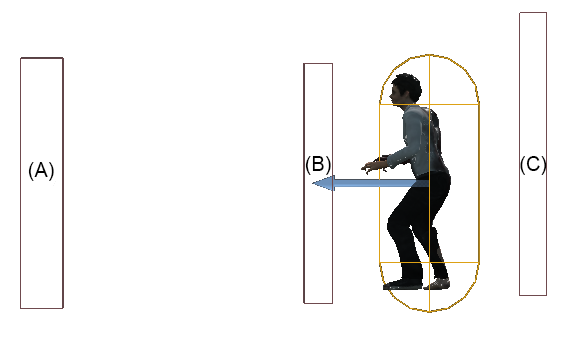}

\caption{A lateral view of the mesh composing the enemy and its associated three intersection bounding boxes, assuming that the avatar approaches the enemy from the left. Box labelled as (\textbf{A}) is used to detect the moment the avatar is close enough to the enemy to trigger an enemy attack. Box labelled as (\textbf{B}) is used for detecting ray--enemy intersections (to determine which object is being attended by the player) as well as bullet-enemy intersections. Box labelled as (\textbf{C}) allow the system to detect the moment the avatar passes beyond the enemy.}
\label{fig:boxesenemy}
\end{figure}

To speed up computation, ray--object intersections (for determining which object is being attended by the player) are tested using pre-computed accessory intersection bounding boxes, rather than using the objects' triangular meshes directly. That is, instead of testing ray intersections against the several triangles present in an object's mesh, intersections are tested against a bounding box properly placed in front of the object. Additional bounding boxes are also associated to all objects present in the scene  to speed-up avatar--object intersection tests. Again, instead of testing intersections between the avatar's and the objects' meshes, intersections are tested between the avatar and two bounding boxes, one~placed in front of the object and another placed behind the object. The former is used to detect the moment the avatar is close enough to the object to trigger a proper interaction (e.g.,~enemy~attack and obstacle dodging) and the latter to detect the moment the avatar passes beyond the object, allowing the object to be removed from the scene. Figures~\ref{fig:boxesobstacles} and \ref{fig:boxesenemy} depict the intersection bounding boxes employed in obstacles and enemies, respectively. The sizes of the intersection bounding boxes had to be enlarged so as to accommodate the inaccuracy of the eye tracker. These adjustments were carried out during a~set of informal tests.

\unskip

\begin{figure}[H]
\centering
\includegraphics[width=12cm]{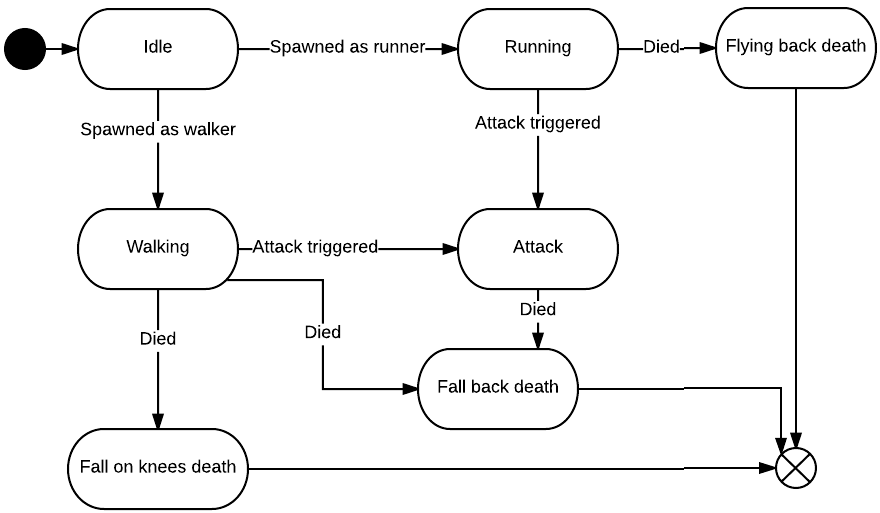}

\caption{The enemy's animation blueprint. The death animation that is triggered from the Walking state is randomly selected to bring variety to the game. The arrows indicate state transitions and the text in them are the conditions necessary to trigger another animation state.}
\label{fig:animblueprintzombie}
\end{figure}

\begin{figure}[H]
\centering
\includegraphics[width=9cm]{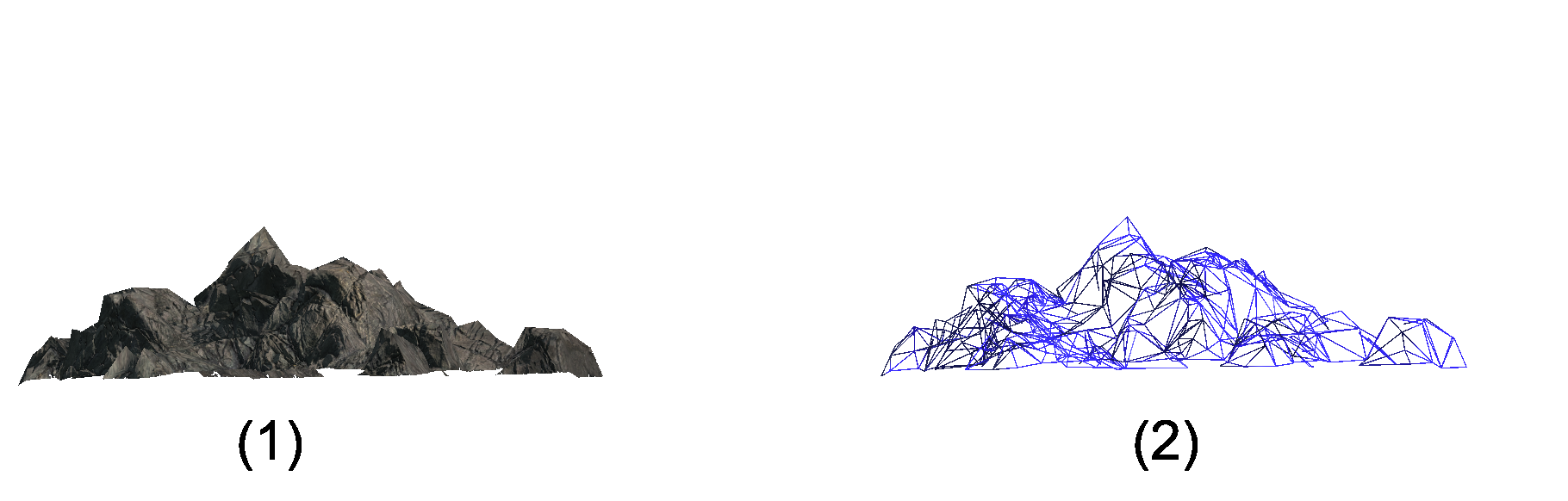}
\includegraphics[width=9cm]{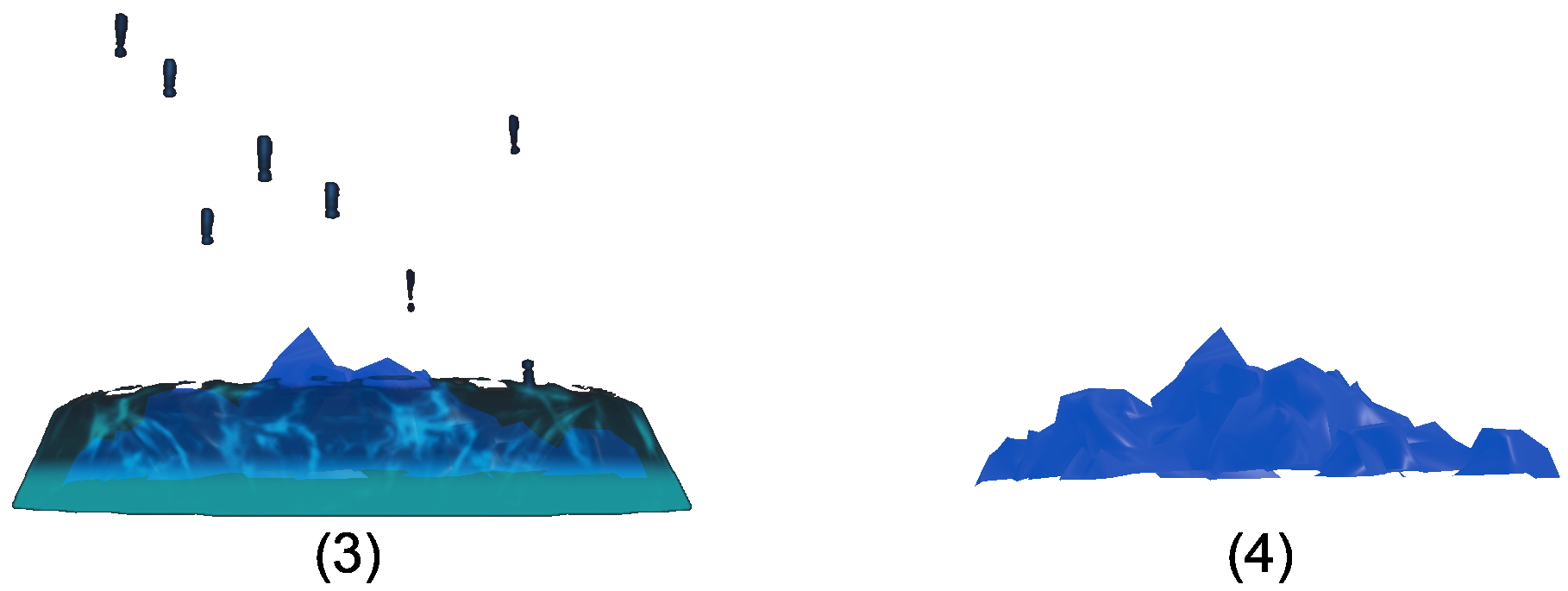}

\caption{The material evolution of a rock obstacle being noticed.}
\label{fig:rocknoticed}
\end{figure}

All assets (e.g., obstacles) were freely obtained from the Unreal's dedicated store. The rigged 3D model and basic animations for the enemies were obtained from Adobe Mixamo. An Unreal Animation Blueprint was built as a state machine which defining the different animation states and transitions between them (see Figure~\ref{fig:animblueprintzombie}). To provide the player with compelling visual feedback, the~appearance of the objects changes when they are first gazed by the player and also later on when they become actually labelled as \textit{noticed}. The first change consists in rendering the object in wireframe with shades of purple. The second change, more dramatic, consists in turning the object in light blue and in adding a particle effect representing a blue shock wave. Figure~\ref{fig:rocknoticed} shows the evolution of the materials and effects used on a rock obstacle through the process of being noticed.

\vspace{-40pt}

\section{Evaluation and Discussion}
\label{chapter:evaluationdiscussion}

As mentioned, several informal tests were run  to guide the development and tune the game's parameters. Then, the whole game was played by a set of ten people in formal game evaluation sessions  to systematically assess what eye tracking technology brings to the game in terms of overall enjoyability.

\subsection{Evaluation Method}

\begin{figure}[H]
\centering
\includegraphics[width=11cm]{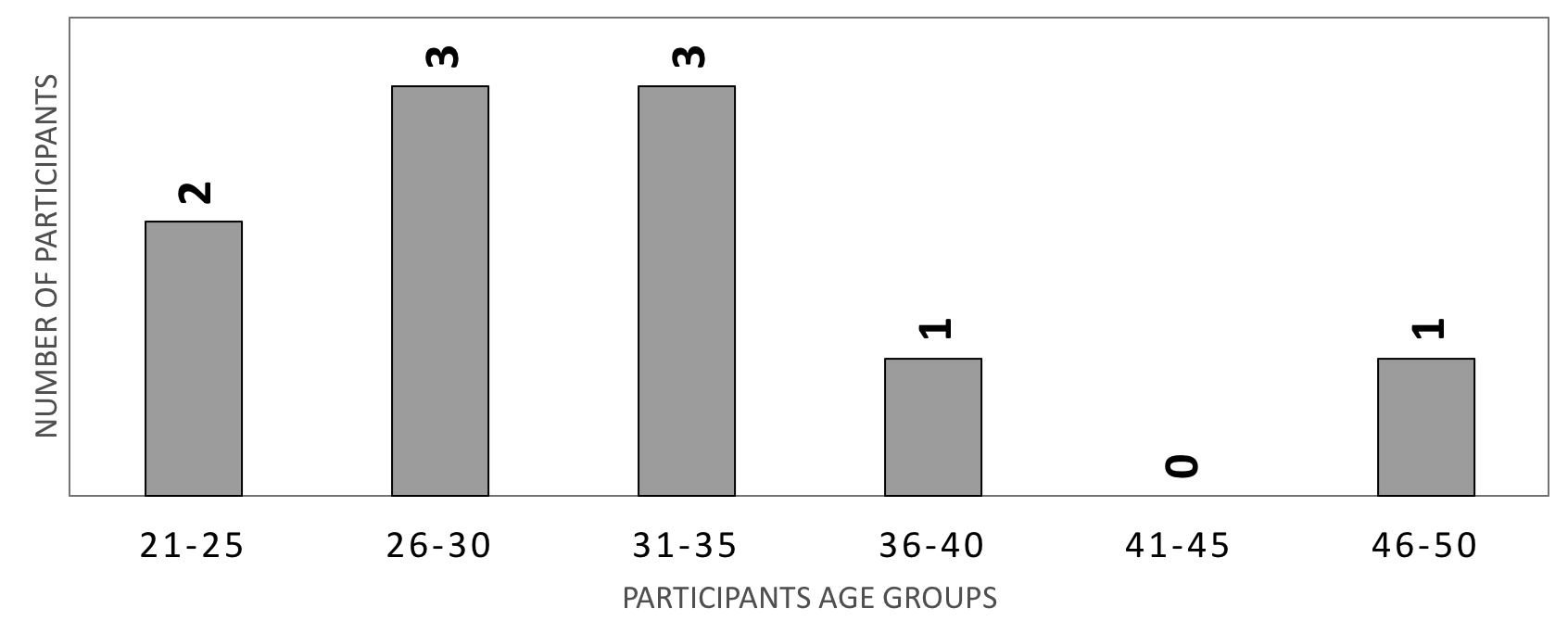}

\caption{Distribution of participants according to their age group.\vspace{-6pt}}
\label{fig:age-distribution}
\end{figure}

Test sessions were carried out privately in a room without the presence of anyone but the participant and the research team. An email-based call for participants that considered themselves gamers was issued to an universe of 170 people. The first ten that responded to the email and complied with the requirements were selected for the test sessions. The participants had no previous knowledge about the game and experience. The ages of the ten participants spanned from 25 to 49~years old (see~Figure~\ref{fig:age-distribution}), with different occupations such as software developer, quality assurance tester, and~student. All participants were male, with no female subjects volunteering for the experience.

In addition to bottom-line data stored during the test sessions, we also enquired the participants with Game Experience Questionnaires (GEQ) \cite{ijsselsteijn_de_kort_poels_2013}, which has been widely applied in previous studies related to control inputs in computer games \cite{drachen_nacke_yannakakis_pedersen_2010,gerling_klauser_niesenhaus_2011}. By using GEQ, we intended to evaluate if the use of eye tracker is enjoyable and comfortable for the player. We also intended to pinpoint possible advantages and disadvantages of the technology in the way it impacts the overall experience of the player and its relationship with the game, comparing the player experience with and without eye~tracking.

\subsubsection{Test Sessions}

\begin{figure}[H]
\centering
\includegraphics[width=15.5cm]{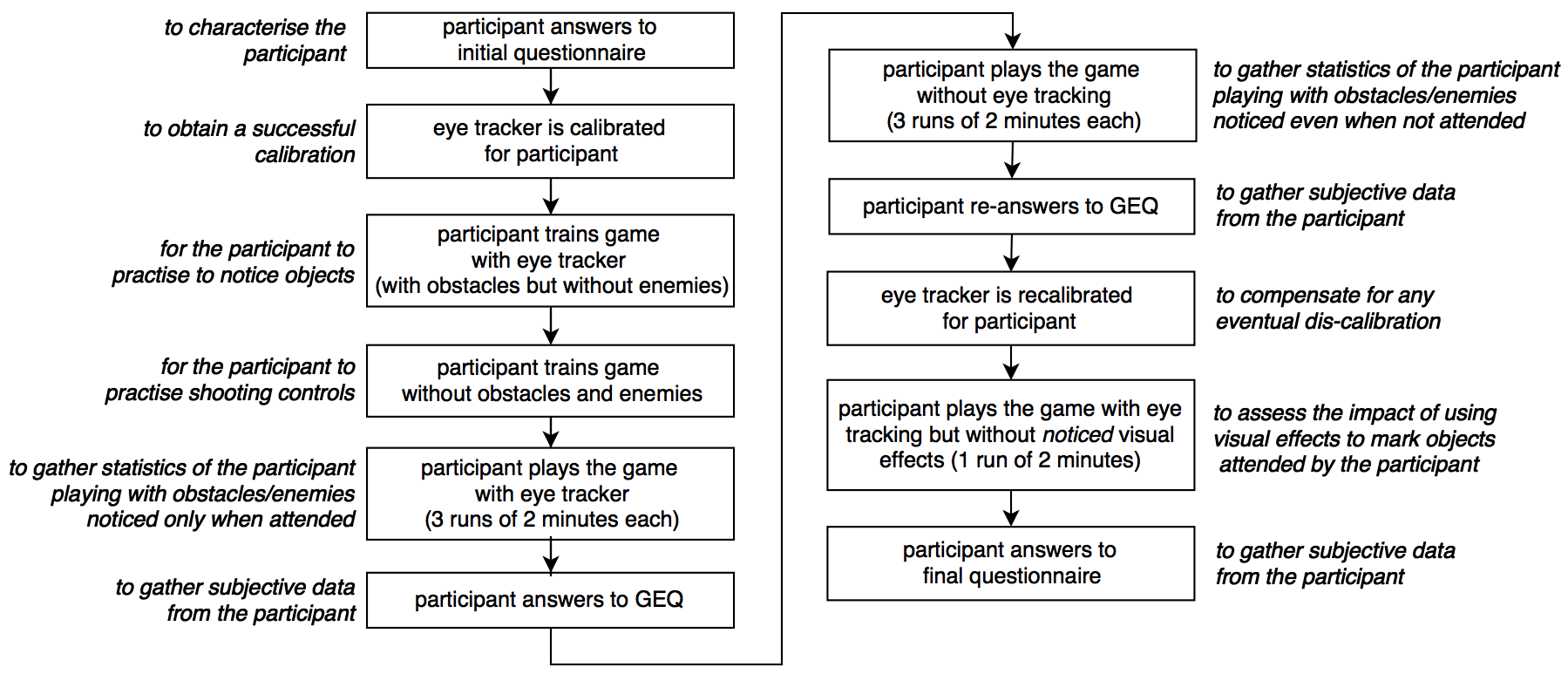}

\caption{Flowchart applied in each test session (per participant). Each box represents a step in the test session and the accompanying off-box italicised text summarises the purpose of applying each~step.}
\label{fig:flowchart}
\end{figure}

\begin{figure}[H]
\centering
\includegraphics[width=13cm]{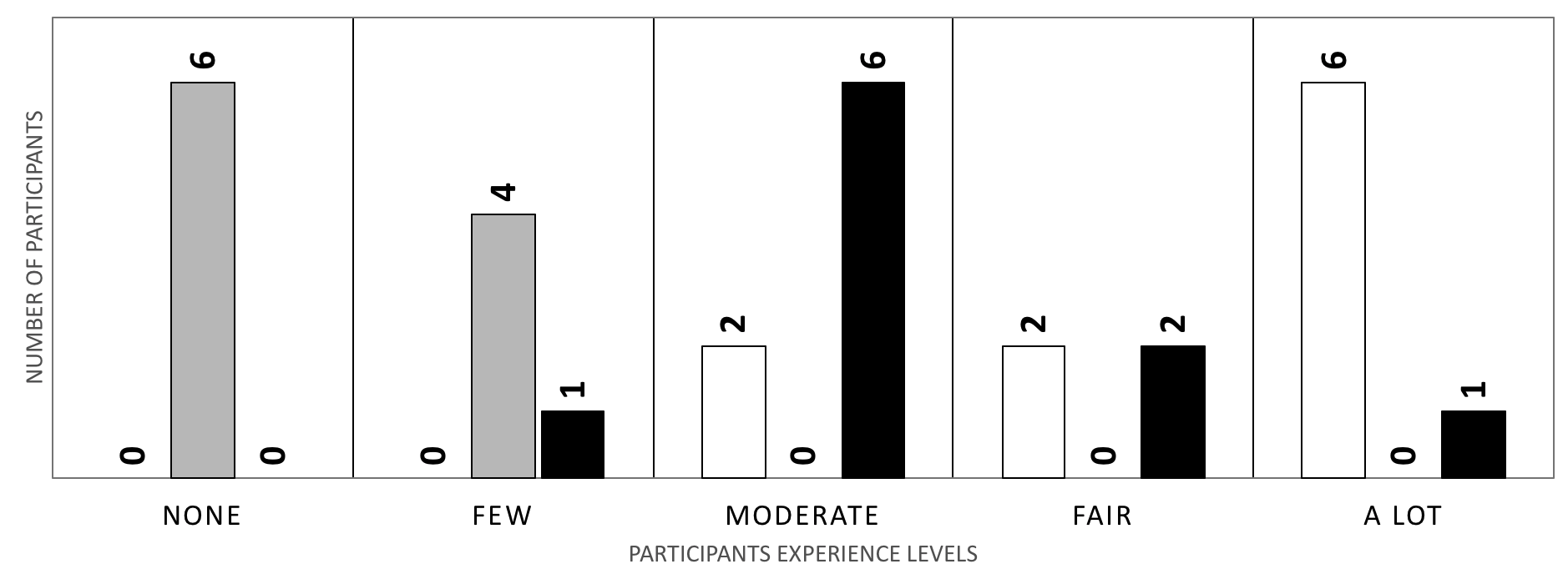}

\caption{Distribution of participants according to their experience with video games (white bars), eye tracking technology (gray bars), and use of gamepads in FPS games (black bars).\vspace{-6pt}}
\label{fig:participants_distribution}
\end{figure}

Figure~\ref{fig:flowchart} presents the flowchart followed in each test session (per participant). Each test session started with a brief questionnaire the participant had to fill in, regarding personal details, such as age and occupation, if the player had some degree of visual impairment, as well as classifying their experience as video game players, with the use of eye trackers, and with the use of gamepads in FPS games. As Figure~\ref{fig:participants_distribution} shows, the average participant had a considerable amount of gaming experience, little to no previous exposure to eye tracking technology, and was moderately experienced using gamepads in FPS games.

\unskip

To reduce any {pleasing biases}, the participants might have, we conveyed the message that we are agnostic regarding the use of eye tracking in games by means of a brief paragraph on the top of the questionnaire with ``the advent of eye tracking technology some developers are including this technology in games'' and ``this study aims at providing scientific validation of the advantages and disadvantages of such inclusion in terms of gameplay and comfort to the user''. Then, a brief explanation of the game's rules and objectives was given, followed by a briefing on the eye tracker calibration process. This process was run as many times as deemed necessary until the calibration was successful (as described in Section~\ref{sec:system-config}). When this successful calibration was achieved, the participants were asked to state their feelings towards the process and if they would imagine themselves doing it at home before playing a game.

To get the participant acquainted with the game, two separate runs were done, with each one only having an input form enabled (gamepad versus eye tracker). These runs had no time limit and did not count for the evaluation. On the first run, only the eye tracker was enabled and no enemies were spawned. The result was a play session with only obstacles being spawned. This allowed the participants to freely use their eyes to notice the obstacles in front of them and get used to this mechanic, without having to worry about the gamepad. The participants were also asked to state any false positives or obstacles that they had noticed but were not tagged as such by the game. On the second run, no obstacles or enemies were spawned. This allowed the participants to get acquainted with aiming and shooting with the gamepad, adapt to its sensibility and button scheme. The main objective was for the participant to feel comfortable with the different inputs and the test session would only advance when the participants confirmed they felt acquainted with the controls.

The participant was then asked to play the game in its original form, as described in Section~\ref{chapter:developmentimplementation}, for~three sessions of 2 min each. For each session, the~ratio between enemies killed and spawned, the~ratio between overall (enemies and obstacles) noticed count and overall spawned count, as well as the number of deaths experienced by the player, were registered. After these three sessions, the player was asked to fill the core and post-game modules of a Game Experience Questionnaire (GEQ)~\cite{ijsselsteijn_de_kort_poels_2013}. This~questionnaire is filled by answering a set of questions whose answers are provided on a Likert-type scale scored from 0 to 4. Afterwards, the participant was asked to play another set of three sessions of 2 min, but this time with eye tracking disabled, which means that all obstacles and enemies were automatically labelled as noticed, with the participant only being required to shoot the enemies. With~these three sessions over, another GEQ was filled in by the player. Then, the calibration process was performed again and the player was asked to play one 2 min session of the game but this time with the visual effects, that occur when the obstacle or enemy is noticed, disabled.

Finally, an end questionnaire was then handed to the participant. This questionnaire was more subjective and consisted of three questions: {(1)} 
``What do you think of the visual effects used on the first sessions, comparing with the last session you played?'' (2) ``How was the overall experience of playing the game?'' (3) ``Would you consider an eye tracker as part of your gaming setup and why?''

\subsection{Experimental Results}

All test sessions were concluded with success, in the sense that all test subjects were able to perform the tasks required from them.

\subsubsection{Eye Tracker Calibration Process}

\begin{figure}[H]
\centering
\includegraphics[width=9cm]{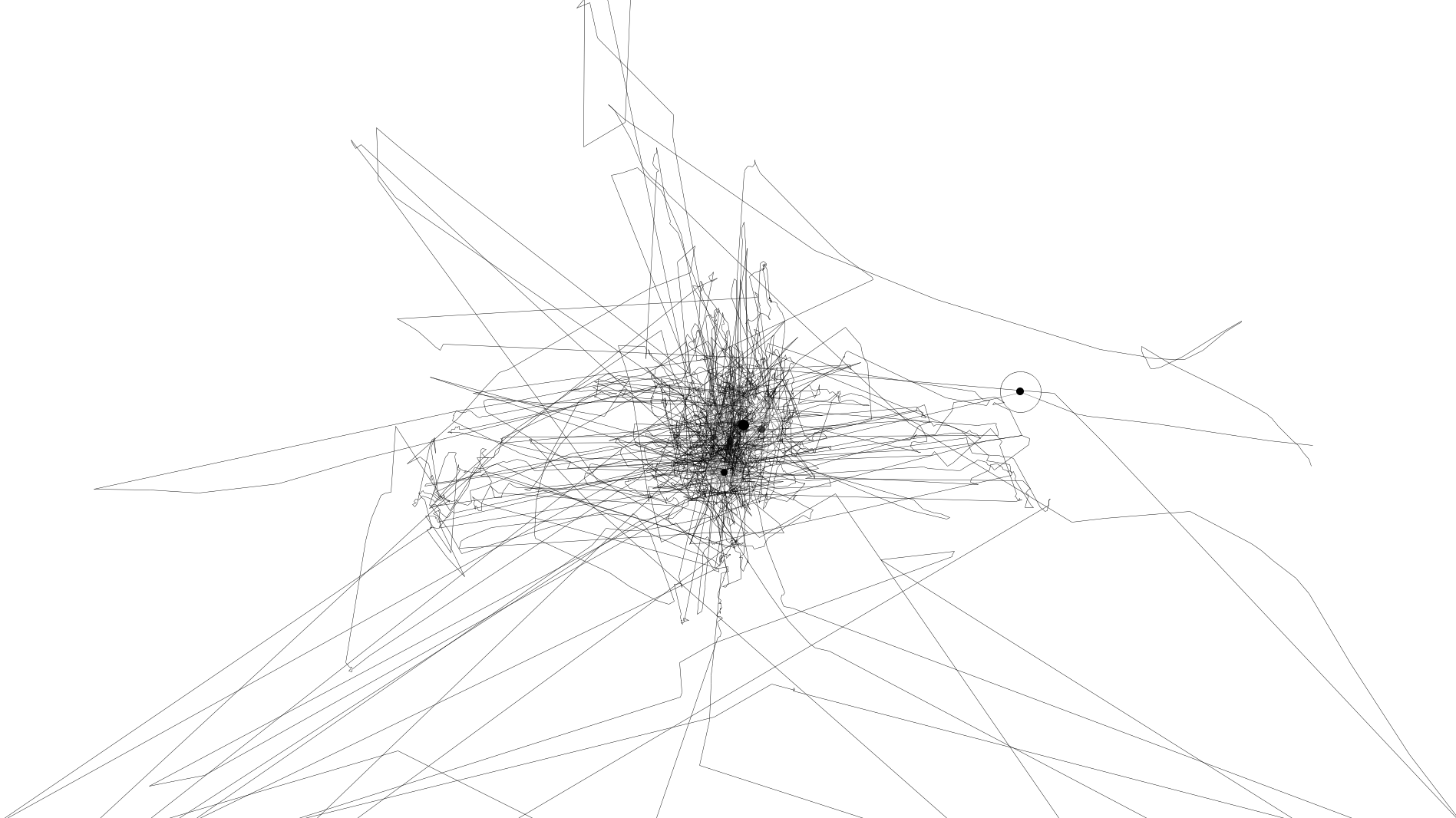}

\caption{Gaze movement of a participant during a typical play-through of Zombie Runner. It can be observed that the player spends most of the time gazing at the central region of the screen.}
\label{fig:eyetrackerpattern}
\end{figure}

\begin{figure}[H]
\centering
\includegraphics[width=9cm]{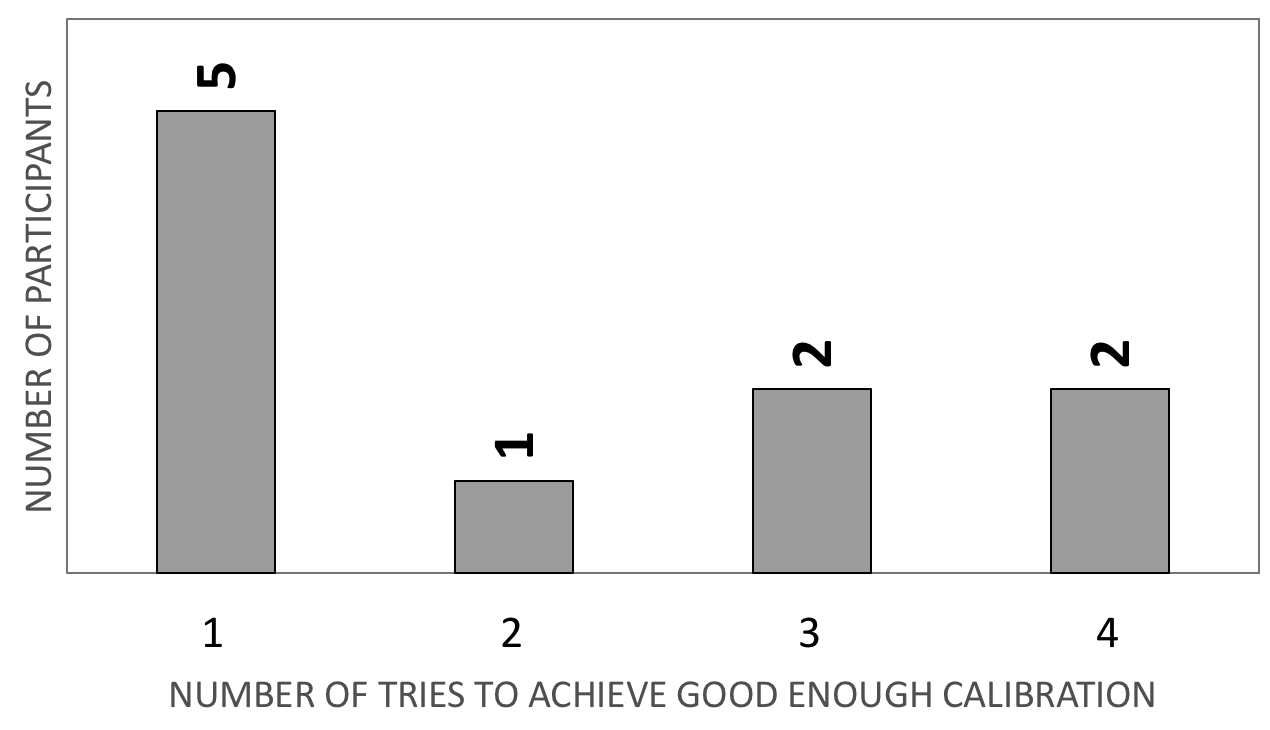}

\caption{Distribution of participants according to the number of eye tracking calibration tries  required to achieve a good enough calibration.\vspace{-6pt}}
\label{fig:triescalibration}
\end{figure}

The calibration process revealed itself as the most challenging step of the testing session. A~calibration was deemed good enough if the accuracy in the areas where the player mostly interacts with in the game, i.e., the whole screen except the top corners (see Figure~\ref{fig:eyetrackerpattern}), was considered as successful according to the criterion defined in Section~\ref{sec:system-config}. Two participants had to remove their glasses so that their gaze could be properly detected. Figure~\ref{fig:triescalibration} shows the distribution of participants according to the number of calibration tries each participant required to obtain a good enough calibration.  As the table shows, a single calibration process was enough for half the participants.

\unskip

After achieving a calibration deemed successful, the participants were asked about their feelings towards it and if they would see themselves repeating this process at home before playing a game. From all participants, eight said that they would see themselves doing it, stating that a calibration process also had to be performed with other controllers. Three of these participants reported that the process was easy and fast, while others stated that it should be easier, but it is still acceptable. The~other two participants did not see themselves calibrating an eye tracker each time they wanted to~play, expressing that they wished the process was easier.

\subsubsection{Play Sessions}

The three 2 min play sessions produced results that suggest an approximation to the ideal game flow, with the participant being able to progressively learn how to use play more competently. This~can be observed in Table~\ref{tab:results_play_session}, which shows the evolution along the three play sessions of: (1) the ratio between the number of killed enemies (by the player) and the total number of spawn enemies; (2)~the ratio between the number of noticed (by the player) elements (obstacles and enemies) and the total number of spawn elements (obstacles and enemies); and (3) the number of avatar (player) deaths. The~two ratios evolved positively, with the player improving in the tasks of killing enemies and noticing them, along with obstacles. The number of deaths went down abruptly from the first to the second session and then went up slightly on the third one, but to a value close to the lowest one obtained in the second session. We suspect this slight increase in deaths can be a consequence of the better results achieved by the participants in terms of killing enemies and noticing objects. To excel in these tasks, participants had to better coordinate the two input forms, which led to a greater risk of being killed.

\begin{table}[H]
\centering
\caption{Results per play session (ratios represented as percentages), provided as STD $\pm$ STE, where~STD stands for standard deviation and STE for standard error of the mean.}
\label{tab:results_play_session}
\begin{tabular}{>{\centering\arraybackslash}p{1.5cm}>{\centering\arraybackslash}p{2.5cm}>{\centering\arraybackslash}p{2.9cm}>{\centering\arraybackslash}p{2.5cm}}
\toprule
\textbf{Play Session} & \textbf{Ratio [\%] of Killed Enemies} & \textbf{Ratio [\%] of Noticed Elements} & \textbf{Number of Deaths}\\ \midrule
1{st} &  $65.6\, \pm 4.4\,$ & $49.7\, \pm 6.3\,$ & $2.3\, \pm 0.2\,$ \\\midrule
2{nd} &  $69.5\, \pm 7.2\,$ & $49.8\, \pm 6.1\,$ & $1.3\, \pm 0.4\,$ \\\midrule
3{rd} &  $79.0\, \pm 5.0\,$ & $52.7\, \pm 5.4\,$ & $1.5\, \pm 0.2\,$ \\\bottomrule
\end{tabular}
\end{table}

The GEQ was filled out by the participants after these three play sessions. As mentioned, after~these sessions, the participants were asked to play an additional set of three play sessions, this~time without using eye tracker, i.e., with all obstacles and enemies automatically labelled as noticed. Table~\ref{tab:averagecomponentcore} compares, for each of the components of GEQ core module (competence, sensory~and imaginary Immersion, flow, tension/annoyance, negative affect, and positive affect), the sessions with and without eye tracking. In the Competence component, the participants felt more competent and able to complete tasks without the eye tracker than with it, which can be explained by the higher degree of difficulty of the game when eye tracking is on. In the Sensory and imaginative immersion component, the participants reported higher scores with eye tracking, which means the feeling of immersion was felt at a greater level with the full game experience. The Flow component was also higher with eye tracking on, which reveals the participants felt a stronger feeling of game flow, a~better balance within the game's Challenge, another component better graded with eye tracking~on. participants reported higher levels on the component of Tension/annoyance with the eye tracker in~use, which~reveals the full game experience can lead to greater levels of frustration. This may be the result of the game being more challenging with eye tracking or of the issues related to the accuracy of the eye tracker (more on this below). Both positive and negative affects are slightly higher when eye tracking was~on, suggesting that players were more emotionally invested in that condition. In both conditions, the~positive affect is considerably higher than the negative one, suggesting that the game provides an~overall positive~experience.

Although GEQ scores are used in this study to compare two designs, we can also analyse the absolute meaning of the obtained scores. In this regard, although the GEQ scores are low when compared to the maximum of the scale, these are not substantially different from the ones obtained for full-fledged commercial FPS games,  reported in a previous study \cite{drachen_nacke_yannakakis_pedersen_2010}. The lower scoring can be explained by the awareness players have of the current state of the art in commercial games. 

\bgroup
\def\arraystretch{1}  
\begin{table}[H]
\centering
\caption{Average values for the different components on the GEQ core module for the play sessions with and without Eye Tracking (ET). Results provided as STD $\pm$ STE, where STD stands for standard deviation and STE for standard error of the mean.}

\begin{tabular}{lcc}
\toprule
& \multicolumn{2}{c}{ \textbf{GEQ Average Score}} \\ \midrule
\textbf{Component} & \textbf{With ET} & \textbf{Without ET} \\ \midrule
Competence & $2.22 \pm 0.27$ & $2.8 \pm 0.25$ \\ \midrule
Immersion & $1.33 \pm 0.3$ & $0.93 \pm 0.3$ \\ \midrule
Flow & $2.08 \pm 0.24$ & $1.68 \pm 0.26$ \\ \midrule
Tension & $1.06 \pm 0.28$ & $0.3 \pm 0.14$ \\ \midrule
Challenge & $1.56 \pm 0.2$ & $0.58 \pm 0.19$ \\ \midrule
Negative affect & $0.68 \pm 0.2$ & $0.45 \pm 0.21$ \\ \midrule
Positive affect & $2.48 \pm0.26$ & $2.26 \pm 0.22$ \\ \bottomrule
\end{tabular}
\label{tab:averagecomponentcore}
\end{table}
\egroup

The results were treated in the same way for the GEQ post-game module, whose results are summarised in Table~\ref{tab:averagecomponentpost}. The differences in levels of positive and negative experience are too small ($\approx$0.05) to allow any analysis from them. The more significant difference in Tiredness and Returning to reality components provide additional support to the idea that eye tracking renders the experience more immersive and engaging. The similar values obtained with and without the eye tracker suggest that the participant's perception of the overall experience was more shaped by the game itself than by the use or lack of the eye tracker. However, these results become more meaningful and easier to understand when compared with the answers the test subjects gave to the last set of informal questions (see below). The main goal of these questions was to extract from the participants more subjective perceptions they had from the game, which could help us understand the GEQ scores.

\bgroup
\def\arraystretch{1}  
\begin{table}[H]
\centering
\caption{Average values for the different components on the GEQ post-game module for the play sessions with and without Eye Tracking (ET). Results provided as STD $\pm$ STE, where STD stands for standard deviation and STE for standard error of the mean.}
\begin{tabular}{lcc}
\toprule
& \multicolumn{2}{c}{ \textbf{GEQ Average Score}} \\ \midrule
\textbf{Component} & \textbf{With ET} & \textbf{Without ET} \\ \midrule
Positive experience & $1.27 \pm 0.31$ & $1.33 \pm 0.28$ \\ \midrule
Negative experience & $0.13  \pm 0.06$ & $0.15 \pm 0.1$ \\ \midrule
Tiredness & $0.5 \pm 0.35$ & $0.23 \pm 0.12$ \\ \midrule
Returning to reality & $0.7 \pm 0.26$ & $0.5 \pm 0.18$ \\ \bottomrule
\end{tabular}
\label{tab:averagecomponentpost}
\end{table}
\egroup

\subsubsection{Informal Questions}

The first question asked participants for their opinion about the used visual effects signalling which objects were labelled as noticed, that is, as an attention feedback mechanism. Nine of the ten participants stated \textit{the importance of the effects as a means to give feedback to the player}, with some pointing out that \textit{without effects the player may be forced to look more than needed as the player never knows if the obstacle or enemy was actually tagged as noticed}. The use of effects was also pointed out as \textit{more rewarding to the players actions}. From all the participants, seven reported that \textit{without the visual effects, the game is more immersive and the experience more realistic, making the way the player looks at things more natural}. This~may suggest that being more immersive does not mean that a game is necessarily more rewarding. In order for it to be both rewarding and immersive, a different, more subtle attention feedback mechanism should be implemented as to avoid breaking the suspension of disbelief. The~development of such a~feedback mechanism (more subtle than the tested visual effects) still demands for additional research. It is also worth studying the value of such an attention feedback mechanism when using highly accurate eye tracking technology. On the one hand, there is the possibility that users stop feeling the need for attention feedback as soon as they feel that the system is accurately tracking their attention. On the other hand, highly accurate eye trackers are expensive and, even those, are unable to fully predict attention deployment, as humans also exploit vision periphery to scan the environment. Some~participants, to whom the eye tracker calibration was not fully successful, stated that \textit{the visual effects used as attention feedback mechanism may induce frustration on them as they could see the discrepancy between where the eye tracker thought they were looking at and where they were actually looking at}. This~issue calls for more robust eye tracking and calibration techniques if we wish to provide universal access to this~technology.

To the second question about the overall experience of playing the game, three out of the ten participants complained about \textit{having to be as static as possible to avoid the eye tracking de-calibration}. Although only two of the participants reported \textit{problems with the game registering when they look at enemies and rock obstacles}, some participants complained about \textit{hardware problems and the frustration of going to the process of calibration and then the technology still not working right}. A participant that had problems with getting the eye tracker to work while wearing glasses \textit{wished that the technology was more prepared for people with glasses}. These issues highlight the biggest practical caveats on the application of eye tracking to video games (and to other related domains): the time-consuming calibration process and brittleness of the system when used in non-ideal scenarios. We expect that further research in eye tracking technology will produce more robust solutions, fostering their use in the wild.  Regarding~the game, a participant stated that \textit{it is well designed}, also expressing \textit{appreciation for the feedback given when the player loses life}. Another participant said that \textit{the mechanic of noticing things is fun, but that the game lacks progression, having nothing new after the first minute}. This might suggest that a more complex game could have been a better fit for this test session, as it would avoid frustration resulting from lack of novelty. In fact, although not so relevant in the context of this study, in which the game sessions were setup to last only a couple of minutes, maintaining player engagement in long testing sessions may demand for more compelling game mechanics. One participant felt that \textit{the time to set an obstacle or enemy as noticed is too long}, which leads to the consider that in future studies these timings should be learned for each player, taking into account the player-dependent uncertainty of the eye tracker. The~experience was \textit{classified as immersive} by three participants, with one of them stating enthusiastically that \textit{eye tracking is amazing}.

With the goal of emphasising potential points of improvement, Table~\ref{tab:negative_comments} provides a compilation of the complaints described in the two previous paragraphs, alongside the number of participants supporting these complaints. As the table highlights, most of the complaints are related to the limitations of current eye tracking affordable technology. The most frequent complaint concerns the lower level of immersion and realism induced by the presence of visual effects used to highlight noticed elements. However, as aforementioned, these visual effects had a positive impact on the overall experience, rendering it more rewarding. Hence,  future studies are required to better analyse this (at least apparent) trade-off between rewarding experience and game immersion/realism in the context of gaze-directed gameplay.

\bgroup
\def\arraystretch{1}  
\begin{table}[H]
\centering
\caption{Number of participants agreeing with a given complaint.}
\begin{tabular}{lc}
\toprule
\textbf{Complaint} & \textbf{Nr. of Participants} \\ \midrule
Impact of noticed-related visual effects to feeling of immersion/realism & 7 \\ \midrule
Need for being as static as possible for proper eye tracking operation & 3 \\ \midrule
Failures in mis-registering enemies/obstacles as noticed & 2 \\ \midrule
Time required for obstacles/enemies to be considered as noticed & 1 \\ \midrule
Problems in tracking the eyes of people wearing glasses & 1 \\ \midrule
Game lacks progression & 1 \\ \bottomrule
\end{tabular}
\label{tab:negative_comments}
\end{table}
\egroup

To the question about whether the participants would consider the inclusion of an eye tracking camera in their gaming setup, the responses were mixed (see Table~\ref{tab:in_the_future}). From the entire group, two of the participants stated  \textit{they would not do it}, with reasons such as \textit{it being another piece of hardware that has little application and would be quickly abandoned after the novelty effect wore off}. Fortunately, we expect eye tracking technology to become embedded in computing devices, reducing the limitations raised by these participants. Three other participants stated that  \textit{they would not in the current state but that they could try it in the future}, stating that \textit{depending on the game it could facilitate precision tasks such as aiming in FPS games or passing the ball to another player in a football game}. This shows that participants see the value of eye tracking as a means to map player's and avatar's attention processes. The reasons these participants presented as to holding off in the adoption of the technology regarded its poor performance while wearing glasses, and the fact that it was not stable or precise, which allied with the calibration process ruined the experience. These issues were already raised by participants in the responses to the two previous questions (please refer to the three previous paragraphs). The other five participants said \textit{they would adopt the technology}, stating that \textit{it was a new form of interaction, that opened new possibilities and created more immersive experiences}.

\bgroup
\def\arraystretch{1}
\begin{table}[H]
\centering
\caption{Distribution of responses to the question ``Would you consider the inclusion of an eye tracking camera in your gaming setup?''. }
\begin{tabular}{lc}
\toprule
\textbf{Response} & \textbf{Nr. of Participants} \\ \midrule
No & 2 \\ \midrule
Maybe, when eye tracking becomes more reliable & 3 \\ \midrule
Yes & 5 \\ \bottomrule
\end{tabular}
\label{tab:in_the_future}
\end{table}
\egroup

These answers showed, along with the previous questionnaires, that the use of eye tracking in games has both pros and cons. On the positive side, the use of gaze-oriented gameplay provided a more immersive and richer experience, providing a better game flow. On the negative side, the~technology's limitations raised feelings in the participants that were not desired. The calibration process and its results, along with some disbelief that eye tracking could have an important role in a computer game, are the main reasons for the participants being adamant about adopting the technology.

\section{Conclusions and Future Work}
\label{chapter:conclusionsfuturework}

This paper presented Zombie Runner, an endless runner FPS game, whose core mechanics and procedurally generated content are modulated by the player's gaze estimated with an affordable eye tracker. A set of testing sessions were carried out  to assess the impact of eye-tracking in the player's satisfaction when playing Zombie Runner. The results obtained from testing sessions show that the use of eye tracking provides a more challenging and immersive experience to the player. These~results complement previous studies, which were mostly focused on performance-based metrics when using eye tracking as a direct control input. Conversely,  Zombie Runner exploits eye tracking as a mechanism to match the avatar's attention model with the one of the player. Moreover, Zombie Runner also exploits eye tracking to guide the procedural content generation of the game's environment, contributing in an original way to the emerging field of Experience-Driven Procedural Content Generation (EDPCG) \cite{yannakakis_togelius_2011}.

During the evaluation, a strong correlation between problems that surfaced with the eye tracker calibration and participants' overall experience was observed. Participants for whom the hardware worked without major flaws reported better levels of satisfaction when contrasting with participants for whom the calibration process was not perfect or took a longer time. Among the participants' complaints about the eye tracking technology, many were related the need to have the head mostly static during the play session, the calibration process requiring repetitions to meet the required accuracy, and overall eye tracker's lack of precision. These complaints show that the affordable eye tracking technology still has to grow and develop until it is in a state that can be accepted by the overall gaming community. However, positive player's experience when calibration was easily attained is a sign that the method will be a valuable asset for the game designer as soon as the eye tracking technology~matures.

As future work, we intend to validate the use of eye tracking in other types of video games and allow for free avatar's movement. We also intend to extend this testing framework, including the philosophy behind the way the player's attention is integrated in the core gameplay, to games with other types of camera perspective, such as third-person games. Finally, we also intend to perform a~more intensive set of tests, enlarging the participant population, to obtain more robust statistics, in~particular to allow an in-depth correlation analysis between gaze patterns, game~progression data, player profiles, and GEQ scores.

\vspace{6pt}

\authorcontributions{J.A. participated in the design of the study, developed the software, participated in the analysis of the results, and participated in the preparation of the manuscript. P.S. participated in the design of the study, participated in the design of the developed software, participated in the analysis of the results, participated~in the preparation of the manuscript, and coordinated all activities.}

\conflictofinterests{The authors declare no conflict of interest.}

\bibliographystyle{mdpi}

\bibliographystyle{mdpi}
\reftitle{References}

\end{document}